\documentclass[preprint]{aastex} % required by pasp

\usepackage{graphicx}
\usepackage{natbib}
\usepackage{url}
\usepackage{color}

\shortauthors{Kleyna, Meech, \& Hainaut}

\date {\centering {\it 17 Sep. 2012; Accepted for publication in PASP}} 

\begin{document}

\title{Faint moving object detection, and the Low Signal-to-Noise
recovery of Main Belt comet P/2008 R1 Garradd\footnote{Based on
observations collected at the Gemini North Observatory (program
GN-20010A-Q-50) and the University of Hawai`i 2.2~m Telescope,
Mauna Kea, Hawai`i, USA, at the European Southern Observatory, La
Silla, Chile (program 184.C-1143), at and the Telescopio Nazionale
Galileo, La Palma, Spain (program AOT20/09B TAC5).}}

%$\rmn{RA}(1950)=19^{\rmn{h}} 22^{\rmn{m}} 18\fs2$,
%$\rmn{Dec.}~(1950)=45\degr 18\arcmin 36\farcs 4$

%%%%%%%%%%%% AUTHORS %%%%%%%%
  \author{J.~Kleyna and   K.~J.~Meech}
  \affil {Institute for Astronomy (IfA), University of Hawai`i, 
          2680 Woodlawn Drive, Honolulu, HI 96822, USA}
  \author{O.~R.~Hainaut}
  \affil{ European Southern Observatory (ESO), Karl Schwarzschild 
     Stra\ss e, 85748 Garching bei M\"unchen, Germany}
%%%%%%%%%%%%%%%%%%%%%%%%%%%%%%

\newcommand{\FIXME}[1] {{\bf \color{red} #1}}
\newcommand{\RAJ}{\ensuremath{\rm RA_{\rm 2000}}}
\newcommand{\DecJ}{\ensuremath{\rm Dec_{\rm 2000}}}
\newcommand{\RAformat}[4]{\ensuremath{#1^{\rm{h}} #2^{\rm{m}} #3\fs #4}}
\newcommand{\Decformat}[3]{\ensuremath{#1\degr #2\arcmin #3\arcsec}}
\newcommand{\DecformatF}[4]{\ensuremath{#1\degr #2\arcmin #3\farcs #4}}

\begin{abstract}

We describe the recovery of faint Main Belt comet P/2008 R1 Garradd
using several telescopes, culminating in a successful low $S/N$
recovery with the Gemini North telescope with GMOS.  This recovery
was a time-critical effort for a mission proposal, and had to be
performed in a crowded field.  We describe techniques and software
tools for eliminating systematic noise artifacts and stellar
residuals, bringing the final detection image statistics close to the
Gaussian ideal for a median image stack, and achieving a detection
sensitivity close to this theoretical optimum. The magnitude of
$R_c$=26.1$\pm$0.2 with an assumed geometric albedo of 0.05
corresponds to a radius of 0.3 km. For ice to have survived in this
object over the age of the solar system, it implies that the object is
a more recent collisional fragment.  We discuss the implications of
the unexpectedly faint magnitude and nuclear size of P/2008 R1 on the
survival of ice inside very small bodies.

\end{abstract}


\section{Introduction} \label{sec:intro}

Main Belt comets (MBCs) are a relatively new class of active
(comet-like) objects in the outer asteroid belt \citep[][for a
  review]{2012AJ....143...66J}.  Although many ideas have been
presented as causes for activity, the only plausible explanation for
those with recurring activity on successive perihelion passages is
water-ice sublimation ({\it e.g.} for 133P/Elst-Pizarro and
238P/Read).  There are two examples of impacts in the main belt which
have produced dust clouds around asteroids (596 Schiela and P/2010 A2
LINEAR), but the most likely cause of activity for the other objects
is a thermally driven process, such as sublimation.

If the activity is driven by sublimation, MBCs represent an unexpected
and persistent reservoir of water in the inner solar system.  The
distribution of water and volatiles in our solar system is a primary
determinant of habitability yet the origin of terrestrial water is
a fundamental unresolved issue in planetary science.  Neither comets
nor asteroids provide an isotopic match to both Earth's water and
noble gas inventory; thus MBCs may play an important role in the
context of understanding the distribution of water in the early
solar system.  MBCs are in near-circular orbits within the asteroid
belt and may contain water frozen out from beyond the primordial
snow line \citep{2008ARA&A..46...57E, 2007ApJ...654..606G} under
different physical conditions than those influencing comets and
asteroids.  Unlike for comets which travel into the inner solar
system, or asteroids which are sampled through their presence in
meteorite collections, the volatiles in MBCs have not been explored,
but they are accessible to space missions.

P/2008 R1 Garradd, discovered in 2008 \citep{2008MPEC....S...46G},
was only the fourth MBC found. The initial observations immediately
following the discovery \citep{JYH09} were at a heliocentric
distance $r=1.8\,\rm AU$ and geocentric distance $\Delta=1.1\,\rm
AU$, with a visible magnitude of $R\approx 20$, after correction
for coma.  This constrained the upper limit to the nucleus
radius to be $R_N < 0.7\,\rm km$, assuming a geometric albedo of
$p_v$=0.05.

Because of favorable trajectories, MBC P/2008 R1 was chosen to be
the primary target of a Discovery class NASA mission called {\sl
Proteus} that was proposed in 2010.  The goal of {\sl Proteus} is
to explore a member of this new class of small bodies in order to
provide insight into the distribution of early solar system volatiles
and thus advance our understanding of solar system habitability.
The mission is a rendezvous mission that is meant to spend 6 months in-situ
to obtain global color maps of the coma and nucleus with $<$ 3-m
resolution, to investigate the bulk physical properties of the MBC,
measure the elemental composition of the dust, and finally, to watch
the onset of outgassing and obtain volatile isotopic and elemental
abundance ratios of several volatiles, including D/H.  Because
P/2008 R1 Garradd had been seen on only one previous apparition,
it was important to recover the object to ensure a high precision
orbit.  Both to obtain astrometry and to characterize the target
nucleus, our group undertook a large effort to recover P/2008 R1.

In late 2009 and early 2010, P/2008 R1 was at heliocentric distances
between $r\approx3.2$--3.5 AU, and geocentric distances $\Delta\approx
2.4$--$3.7$ AU, implying that P/2008 R1 would be 3 to 4 magnitudes
fainter than during its discovery observations, assuming the usual
$r^{-2}\Delta^{-2}$ geometric flux dependence. At these distances
the object was expected to be inactive, making the observations very
challenging.  

In this paper, we describe the P/2008 R1 recovery observations of
2009 to 2010, culminating in a successful low $S/N$ recovery.  In
addition to the observations presented below, the MBC field was
imaged for 90 minutes with the 10~m Keck I LRIS instrument on Mauna Kea
on 2010 Mar. 17 in moderately good seeing, ending in a non-detection
(Jewitt, private communication).

We present an optimized method of shifting and adding astronomical
images to recover moving objects at the detection threshold of the
instrument.  We also describe the construction of a detection pipeline
using readily available software normally designed for static images.
This pipeline has the advantages of fitting astrometric coordinates,
of producing an accurate astrometric measurement as a final product,
and of allowing masking operations that suppress stellar residuals and
allow a detection sensitivity that approaches the theoretical optimum
for median stacked images.  Finally, we discuss the implications of
P/2008 R1's unexpectedly faint magnitude and small size for the
persistence of ice in the asteroid belt.

\subsection{Archive searches}

We searched the main astronomical data archives for pre-discovery
serendipitous images of P/2008 R1. In the ESO Science
Archive\footnote{http://archive.eso.org}, we checked every image
acquired between January 1996 and June 2010 with an exposure time
longer than 10~s using the ESO Wide Field Imager (WFI, 2.2m), the ESO
Faint Object Spectrograph and Camera 2 (EFOSC2), SofI (3.6m), Wide
Field Infra--Red Camera (WFCAM, 4m), Visible Multi--Object Spectrograph
(VIMOS), Focal Reducer and Low Dispersion Spectrographs 1 and 2 (FORS1
and FORS2), HAWK-I (High Acuity, Wide--Field K--Band Imaging), and NaCo
(Nasmyth Adaptive Optics System [NAOS], Near--Infrared Imager and
Spectrograph [CONICA], 8.2m) against the ephemerides of the comet for
the corresponding instrument's field of view. The only frames returned
that matched the ephemeris position were the NTT images described
below (Section~\ref{sec:NTT}).

Using the SSOS
tool\footnote{http://www2.cadc-ccda.hia-iha.nrc-cnrc.gc.ca/ssos/}, we
also searched for images from MegaCam on CFHT (3.6m), SuprimeCam
on Subaru (8.3m),GMOS on Gemini (North and South; 8.2m) and WFPC2,
ACS, and WFC3 on HST (2.5m in Earth orbit) from 1994 until 2010 June
10. The only frames reported were those described in Section
\ref{sec:GN}.

\section{Observations}  \label{sec:obs}

Between November 2009 and April 2010, we undertook an extensive
observing program using several telescopes, ranging in aperture
from 2.2m to 8m.  These are briefly described below and summarized
in Table \ref{tab:observations}.

\subsection{Telescopio Nazionale Galileo}\label{sec:TNG}

CCD images were obtained on the 3.56m Telescopio Nazionale Galileo
(TNG) on La Palma on the nights of UT 2009 Nov. 19 and 20. The
DOLORES\footnote{http://www.tng.iac.es/instruments/lrs} instrument
was used with the Johnson-Cousins $R$ filter R\_John\_12. The pixels
were $0\farcs 25$ on the sky, resulting in a field of view (FOV) of
$8\farcm 6$ on the E2V 4240 CCD. The data were obtained at low to
moderate airmasses (1.1$<$$\chi$$<$1.9) with the seeing ranging
between $1\farcs 2$ to $1\farcs 8$.

\subsection{New Technology Telescope}\label{sec:NTT}
The observations were performed on the ESO 3.56m New Technology
Telescope (NTT) on La Silla, with the ESO Faint Object Spectrograph
and Camera (v.2) instrument
\citep[EFOSC2]{1984Msngr..38....9B,2008Msngr.132...18S}, through a
Bessel $R$ filter, using the ESO\#40 detector, a 2k$\times$2k
thinned, UV-flooded Loral/Lesser CCD, which was read in a 2$\times$2
bin mode resulting in $0\farcs 24$ pixels, and a $4\farcm 1$ FOV.
The data were acquired on UT 2010 January 13, under dark
and photometric conditions. However, because of the high declination
of the object, the airmass ranged from 2.6$<$$\chi$$<$3.0, resulting in a
poor on-chip image quality of 1$\farcs$5 FWHM.

\subsection{University of Hawai`i 2.2m}\label{sec:UH88}
\begin{figure}
%\raisebox{7cm}{\bf a.}\includegraphics[width=8cm]{fig_TNG_2009-nov.ps}
%\raisebox{7cm}{\bf b.}\includegraphics[width=8cm]{fig_NTT_2010-jan.ps} \\
%\raisebox{7cm}{\bf c.}\includegraphics[width=8cm]{fig_UH_2010feb.ps}
\includegraphics[angle=0,scale=0.8]{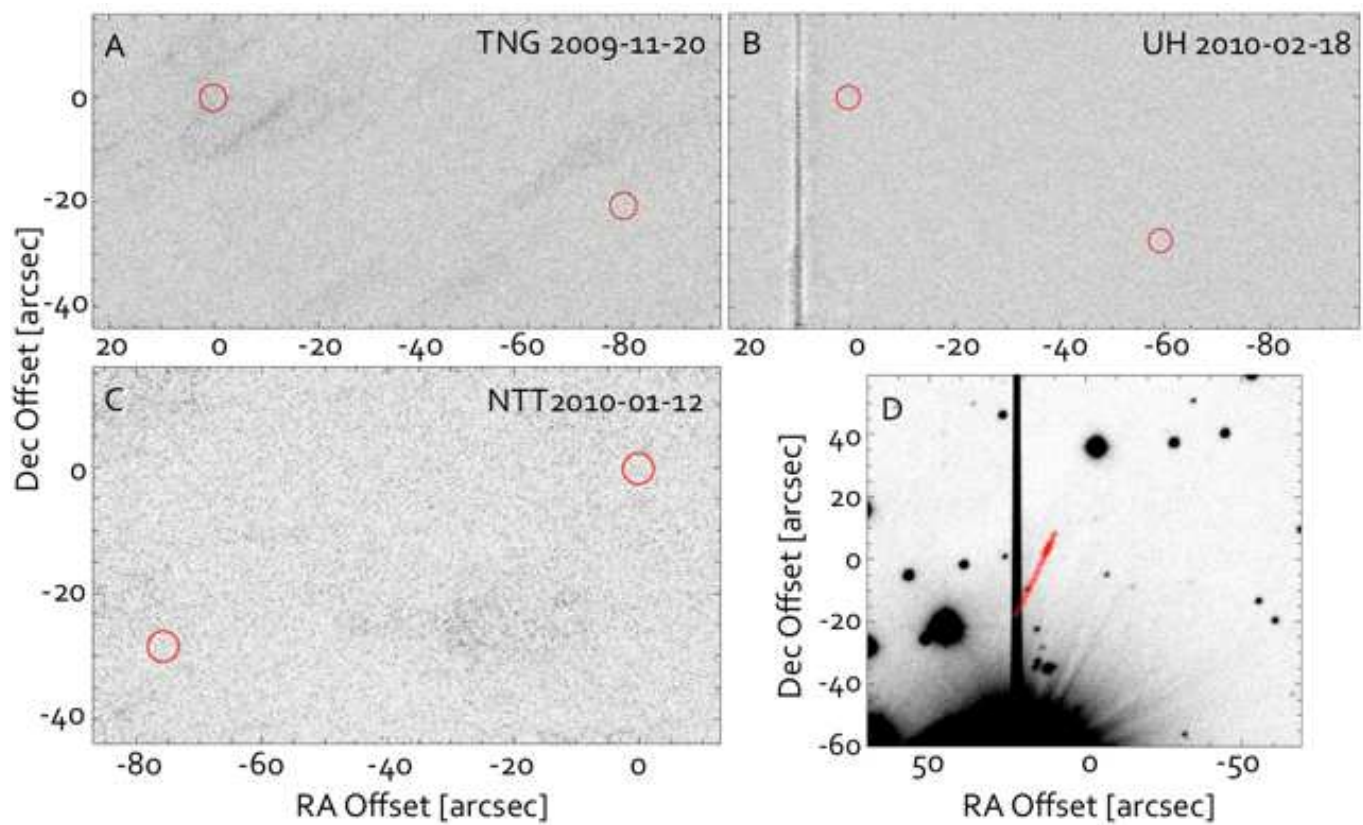}
\caption{\label{fig:image-sec}
Sub-frames of composite images centered on the expected position of
P/2008~R1 Garradd, in  
{\bf A.} with the TNG, 
{\bf B.} with the UH~2.2m, and 
{\bf C.} with the NTT.
Panels A, B and C show the median of the star subtracted frames.
The red circles at RA and Dec offsets (0,0) show the expected
ephemeris position, and the second dot shows the position of the
MBC using the new ephemeris based on the recovery.  Panel D shows
the UH~2.2m data stacked as a composite image without background
star removal, using all frames where the expected position of P/2008
R1 was not severely affected by the charge bleed from the field
star.  frames. See Table \ref{tab:observations} for details of the
observations.
}
\end{figure}

The observations were performed on the University of Hawai`i (UH)
2.2-m telescope on Mauna-Kea, using the Tektronix 2k$\times$2k CCD
camera with the Kron-Cousin $R$
filter\footnote{http://www.ifa.hawaii.edu/88 inch}. The projected
pixel size is $0\farcs 219$, yielding a FOV of $7\farcm 5$. The
exposures were obtained, under dark and photometric conditions.
High winds degraded the seeing to worse than average at 1$\farcs$1 FWHM.
A nearby very bright star significantly contaminated the expected position
of the comet (Fig.~\ref{fig:image-sec}D).

\subsection{Gemini North}\label{sec:GN}
\begin{figure}
\includegraphics[angle=-90,scale=0.8]{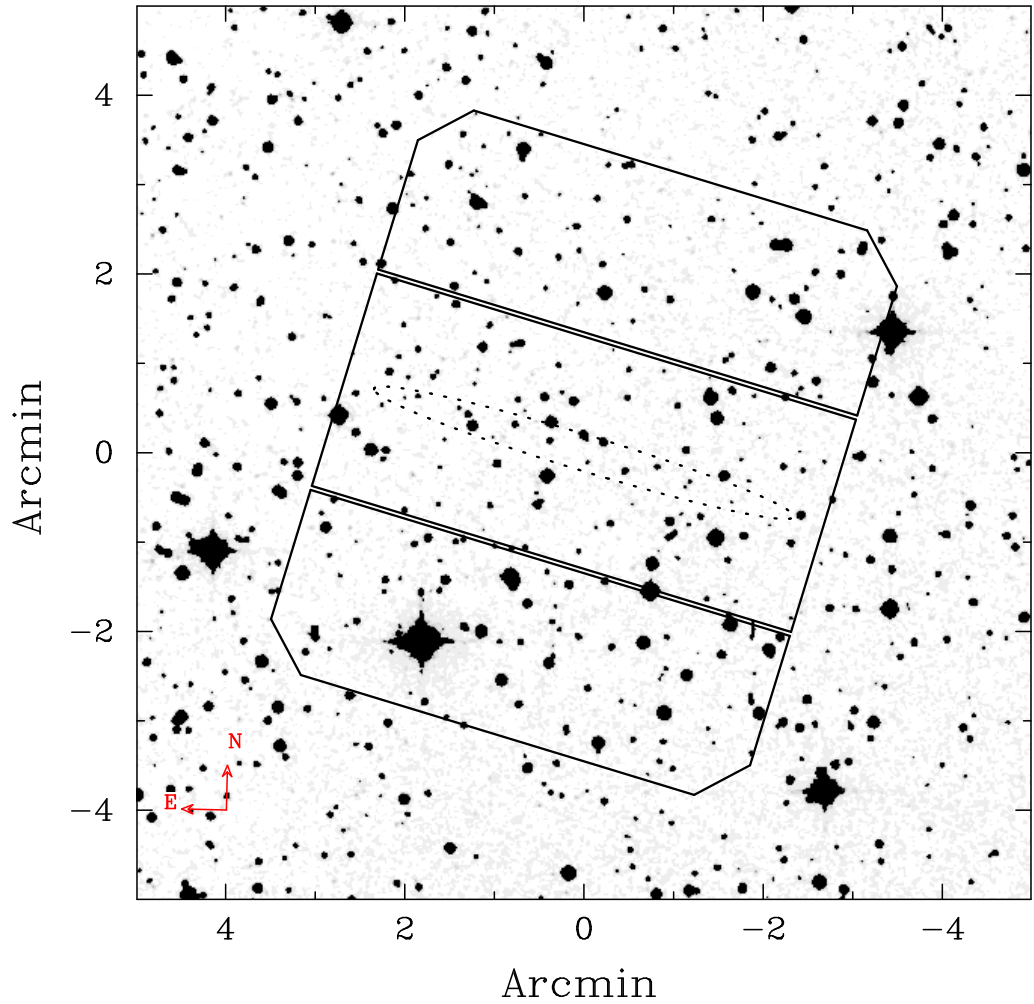}
\caption{\label{fig:finder}
Finder chart for MBC P/2008 R1, centered on
\RAJ$=$\RAformat{05}{55}{30}{3}, \DecJ$=$\Decformat{+35}{14}{16},
showing the 99\% uncertainty ellipse and the GMOS field orientation
(at PA = $73^\circ$), designed to place the long axis of the
uncertainty ellipse along the unvignetted center chip.
}
\end{figure}

Data were collected on UT 2010 April 4 using the Gemini North
telescope GMOS instrument
\citep{2004PASP..116..425H,2002PASP..114..892A} in imaging mode.  GMOS
is a $5.5\arcmin\times 5.5\arcmin$ FOV instrument with
three $1056\times2304$ (binned) detector arrays, two of which are
vignetted.  At the $2\times2$ binning used, the pixel scale is
$0\farcs145$.  As shown in Fig.~\ref{fig:finder}, the field was
rotated so that the uncertainty ellipse lay along the long axis of the
center chip.

Data were acquired under dark photometric conditions through the
$r^\prime$ band (GMOS filter ID G0303) with seeing around $0\farcs5$,
and an airmass between 1.35$<$$\chi$$<$2.2.  Non-sidereal guiding was used,
at a rate of ${\rm 32\arcsec \, hr^{-1}}$, so that each star was
trailed to a length of $2\farcs7$.  Combined with the relatively
densely populated field (Galactic longitude $\rm \ell_{II}=175$, and
latitude $\rm b_{II}=5$), this trailing tended to fill a large fraction
of the field with stars.

\section{Data Processing} \label{sec:proc}

\subsection{Traditional Reduction Method: TNG, NTT and UH}

The data from the TNG, UH and the NTT were processed using customary
methods: after bias subtraction and flat-field correction,
the individual images were aligned using the background stars as
reference, and were co-added, forming the the star background
template. This template was astrometrically calibrated, scaled to
the appropriate exposure time and subtracted from the individual
frames, which were then shifted to account for the motion of the
comet (using the Horizon ephemerides and the astrometric calibration).
The individual, star-subtracted frames were averaged with a median-like
rejection algorithm. The method is described in more detail in
\cite{Hai+04}.

Figure~\ref{fig:image-sec} displays subsets of the frames centered
on the expected position of the comet for the data from the TNG,
the NTT and the UH2.2m. None of the images show an MBC candidate
at the expected ephemeris position.  In order to check for very
faint candidates, the individual frames were combined using various
subsets of the images, and the star-subtracted stacks were compared
with the stacks including the stars. No convincing candidate in any
of the subsets was identified.

The limiting magnitudes were estimated by inserting scaled PSFs into
the images. The adopted limiting magnitude is that for which at least
65\% of the fake comets are visible. The values are listed in
Table~\ref{tab:observations}.  With the exception of the NTT
observations, the limiting magnitudes are considerably fainter than
the value reported from the observations by \citet{JYH09},
suggesting correspondingly smaller nuclei.

\subsection{New Reduction Techniques for the Deeper GMOS/Gemini Data}

For the more crowded Gemini fields, we constructed a pipeline from the
TERAPIX/ASTROMATIC tools of Emmanuel
Bertin\footnote{\url{http://www.astromatic.net}}, which provides a
more automated and flexible way of performing the tasks of aligning
the images, calibrating the astrometric solution, and subtracting
background. This allowed us to optimize the subtraction and reach the best
limiting magnitude. Henceforth all software commands mentioned will be
part of TERAPIX/ASTROMATIC unless otherwise noted.

The first step utilized the {\sc SExtractor} photometry program to
create a catalog of stars for each GMOS image.  Next, we employed the
{\sc scamp} astrometric program to fit a linear world coordinate
system (WCS) to each image, using the {\sc sextractor} catalogs as an
input and 2MASS as the astrometric reference catalog.  A typical fit
employed $\gtrsim 80$ stars with an RMS of $\lesssim0\farcs25$ in each
dimension. To shift and add the night's observations, we modified the
center RA and Dec of the WCS output files produced by {\sc scamp} in
the reverse of the object's motion, using the observation time in the
header.

Finally, we performed the co-addition using the {\sc SWarp} program,
which background--subtracts and rebins the images to a common
astrometric frame, and performs a median.  Images were
rebinned to a $0\farcs2$ pixel scale, with conventional North--up,
East--left linear world coordinates.

This simple procedure failed to show the comet, because stellar
residuals overwhelmed any signal from the MBC P/2008~R1.  Thus, we
next created an un-shifted median deep stack of all of the images
using {\sc swarp}.  We then subtracted this deep stack from each
individual exposure by minimizing the sum of the absolute value of the
output image, a criterion designed to minimize the amount of final
flux present, adjusting for the changes of extinction and seeing
during the observations. 

Using the median-subtracted image for the shifted stack substantially
reduced the background artifacts, but remaining traces of stars still
produced too many low level features.  These stellar residuals appear
to result from a combination of slight PSF variations through the
night, as well as possible small astrometry errors. For instance,
extinction and seeing may have changed during an exposure, so that the
centroid of the trailed stars may have shifted,
resulting in slight errors in the final WCS. After subtraction, the maximum
flux in the stellar residuals is about 5\% of the flux of the star.

As the final step of cleaning the individual images, we created a mask
image for each median-subtracted exposure.  Whenever a pixel in the
exposure exceeded 500 photons, or whenever a pixel in the median
exceeded 100 photons, the weight mask was set to zero.  The typical
masked fraction was about 11\%.  {\sc swarp} then ignored
zero-weighted pixels when producing the final shifted stack.  This
masking process yielded a faint detection of P/2008 R1.

\begin{figure}
\begin{center}
\vskip -6cm % for some reason this figure is placed low
\includegraphics[angle=0,scale=0.8]{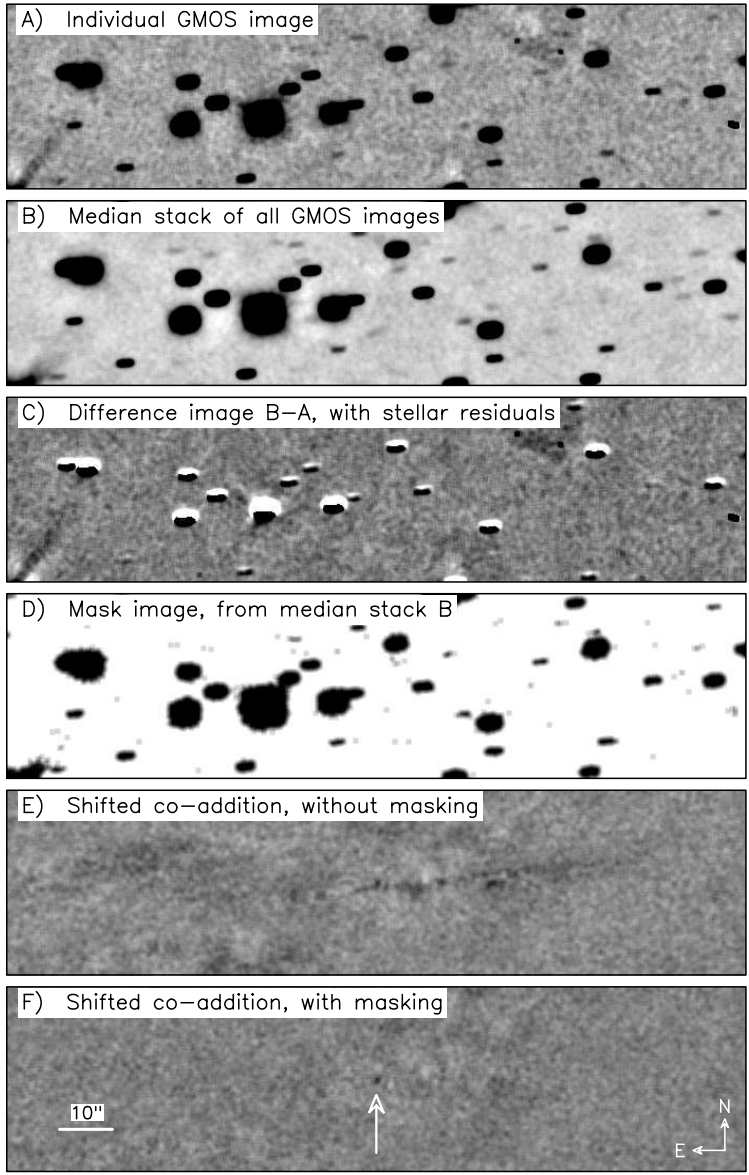}
\end{center}
\caption{\label{fig:image-tech}
Image sections showing the reduction steps for the recovery. A: The
first image of the run; B: The median of all of the images, projected
onto the coordinates of A; C: The difference between the median
stack and the first image, using scaling from automated stellar
photometry; D: A mask image where dark pixels are those pixels in
C that are to be ignored, created by noting pixels in both the stack
B and individual image A that are above respective thresholds; E:
The final detection stack, omitting the masks in D; F: The final
detection stack, co-added at the rate of motion of P/2008 R1. The
detection of P/2008 R1 is indicated by the arrow.
}
\end{figure}

Fig.~\ref{fig:image-tech} shows the sequence of images in the reduction
procedure as described above. In panel E of the figure, the un-masked
stack, MBC  P/2008 R1 is visible but is connected to an extended artifact,
so it cannot be identified without reference to the masked stack
in panel F.  Additionally, we performed the usual verifications to
ensure the reality of the detected object.  Image stacks were created
from subsets from the full frame set; they also show the comet.
Individual frames were checked for artifacts and/or contamination
at the position of the comet; nothing was found. This convinces us
of the reality of the object.

The astrometry (Table \ref{tab:astrometry}) of the image shows that P/2008 R1 was found $\sim51\arcsec$
west and $\sim16\arcsec$ south of the expected ephemeris position on
2010 April 4, but was along the error ellipse and was reported in
\cite{MPC70591}.  With the updated orbit we again searched the TNG,
NTT and UH CCD frames and the comet was not found.  In the TNG and
NTT data this was because the limiting magnitudes were not deep
enough.  For the UH2.2m data, the full image stack probably had a
sufficiently deep limiting magnitude, but the position of the MBC
was located under the charge bleed from the bright star for some
fraction of the time, and the partial stack was not deep enough.

\subsection{Detection significance}

Aperture photometry of the final median stack gives a flux of
$1652\pm356$ photons per 300 seconds in a $r=5$ pixel ($1\arcsec$)
radius aperture, corresponding to a magnitude of $r^\prime=26.28\pm
0.2$, using the published GMOS calibration\footnote{GMOS North
  $r^\prime$ magnitude from flux: $r^\prime=28.20-2.5\log_{10} ({\rm
    photons\,s^{-1}}) - 0.11\times ({\rm Airmass} -1)$, from
  \cite{jorgensenGMOSCalib} and the GMOS calibration web page.}.  A
Moffat function fit over a $4\farcs2$ square gives a similar flux of
1594 photons ($r^\prime=26.43$), while a Gaussian fit, which tends to
omit the PSF wings, gives a flux of 1218 ($r^\prime=26.61$).  Both
fits yield a FWHM of $0\farcs6$, suggesting that the frames are well
aligned.  We estimate the significance of the detection by dividing
the flux of the detection by the standard deviation of the flux in
randomly placed apertures.  The significance is $6.0\sigma$ for a 3
pixel radius aperture, $4.6\sigma$ for a radius of 5 pixels, and
$3\sigma$ for a radius of 7 pixels.  We therefore report the magnitude
of the comet as $r^\prime=26.3 \pm 0.2$.  Although one might expect
the extended Moffat flux to be larger rather than smaller than the
aperture flux, the difference is smaller than the photometric
uncertainty, and is likely to be influenced by random variations in
the Moffat wings.

After rescaling the exposure times to account for the 11\% masking
fraction, the GMOS exposure time calculator (ETC) predicts a
$4\sigma$ to $5\sigma$ detection within a $1\arcsec$ pixel radius aperture,
depending on observing conditions.  We conclude that this technique
can recover objects very close to the theoretical limit, even in a
moderately crowded field.

\begin{figure}
\begin{center}
\includegraphics[angle=-90,scale=0.8]{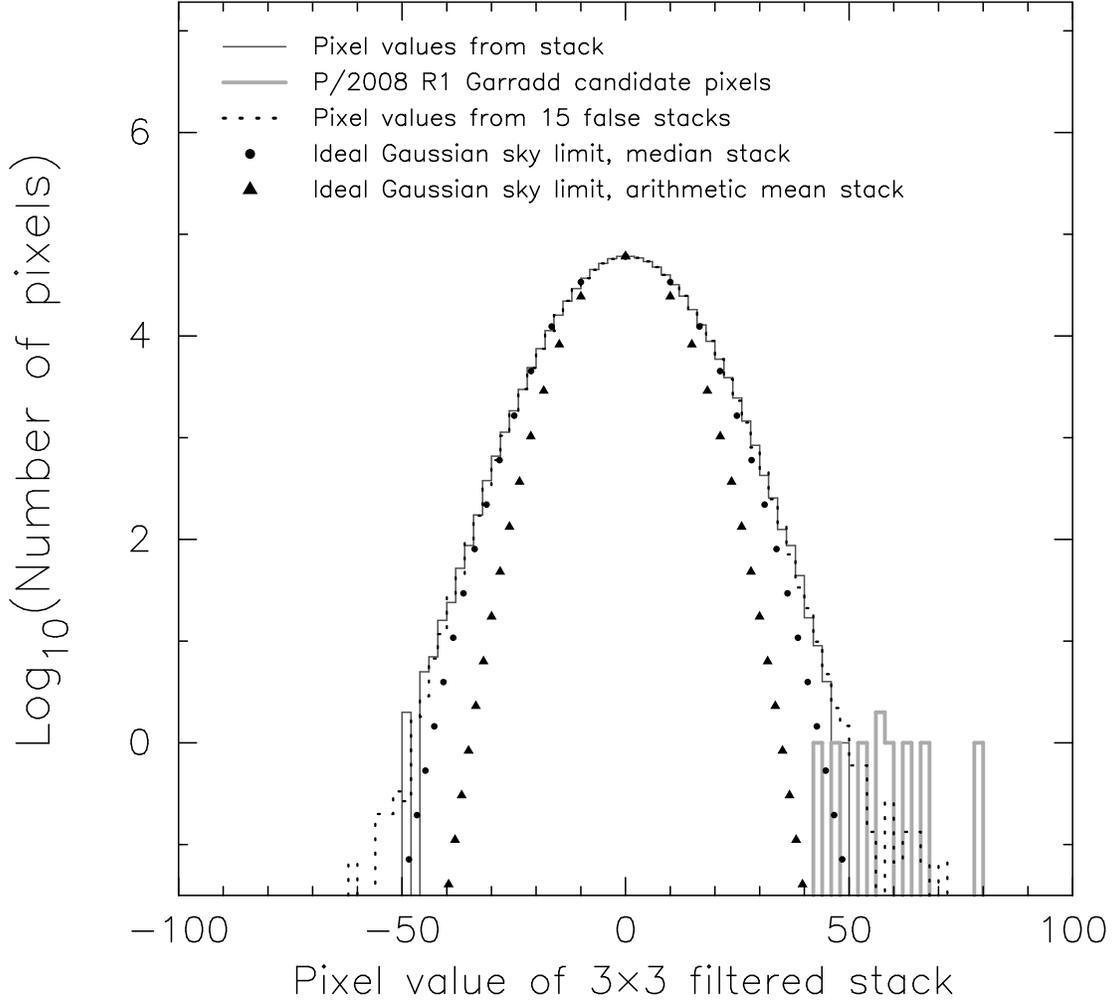}
\end{center}
\caption{\label{fig:pixhist} Logarithmic histogram of the central
  $900\times800$ pixels of the detection stack after $3\times 3$
  boxcar smoothing (thin line); the 9 center pixels of the object are
  depicted by the thick gray line.  The dotted line is the normalized
  histogram of 15 null detection stacks created like the correct
  stack, but with a random $2\arcsec$ jitter.  The object was detected
  only because the overall image statistics are nearly Gaussian
  (parabolic in the logarithm) because of the careful subtraction and
  masking procedure. The triangles show the expected pixel
  distribution if the only noise contribution were Gaussian sky
  counts, assuming arithmetic mean stacking. The circular dots are a
  similar optimal limit for median stacking, rather than arithmetic
  mean.}
\end{figure}

Fig.~\ref{fig:pixhist} shows a histogram of the pixel values of a
$3\times3$ smoothed version of the final detection stack, as well as
pixels drawn from false stacks that have had a deliberate $2\arcsec$
random jitter introduced during shifting to eliminate the moving
object.  Despite the object's low detection significance, it still
contains the most outlying pixels of the stack image, eclipsing any
detections in the false stacks.  From the parabolic shape of the
histogram, it is evident that the final pixel distribution has a
minimal non--Gaussian tail.     

We compare the pixel distribution in the stack to the ideal Gaussian
sky limited case, using the empirical $\pm 1\sigma$ pixel flux
distribution in the 20 background subtracted component images as a
measure of image noise.  For the ideal cases, we assume a zero masking
fraction, in order to ascertain losses associated with masking.  The
solid triangles in Fig.~\ref{fig:pixhist} show the expected final
pixel distribution for simple arithmetic mean stacking with pure
Gaussian sky noise.  The solid circular dots show the effect of median
stacking, which increases the spread by a factor of 1.25.  The fact
that the observed pixel histograms closely track the circular (median)
dots shows that this method approaches the sensitivity of the
median-stacked, purely Gaussian noise dominated ideal case.  The 11\%
masking fraction is responsible for only a modest loss of sensitivity.

\section{The Unexpectedly Small Size of P/2008 R1}

To allow comparison of our observations with customary photometric systems, we
convert our magnitude $r^{\prime}=26.3\pm0.2$ to Kron-Cousins $R_c$ band
using the color equations given by \citet{Fukugita+96}: $R_c =
r^\prime -0.16 (V-R) -0.13$. \citet{JYH09} reported $V-R = 0.37\pm0.05$
for the active comet, and \citet{HBP12} list $V-R = 0.49\pm0.10$
as the average of 49 Short Period Comet nuclei; the uncertainty in
$V-R$ is much less important than the observational error in 
$r^\prime$. Using $V-R=0.4$, we get $R_c = 26.1 \pm 0.2$. Given the
helio- and geocentric distances and solar phase angle listed in
Table~\ref{tab:observations}, and using the same $G=0.15$ slope parameter
as \cite{JYH09}, this results in an absolute magnitude $H_R = 19.6
\pm 0.2$. Using a geometric albedo $p=0.05$, this corresponds to a
radius of $R_N=0.3$~km, where the radius uncertainty is about $\pm 10\%$
from the magnitude error alone.

From the value of $R_C$, we remark that the limiting magnitudes we
reached with the other telescopes were indeed too shallow.

The radius we obtain is consistent with the limit proposed by
\cite{JYH09} based on measurements of the active comet ($r<0.7$
km), and is even smaller than their conservative rough estimate ($r\sim
0.5$ km).  \cite{Sarid12} have undertaken a parameter study of the
survival of ice in MBCs, and have found that the only volatile able
to survive the age of the solar system in the orbital parameter space
occupied by the MBCs is crystalline water ice.  Even then, the
survivability is dependent upon the average orbital radius, the MBC
radius and its density. With an average orbital radius $\rm
a_c=a(1-e^2)$ = 2.4 AU where $a$ is the semi-major axis [AU] and
$e$ the eccentricity, it is unlikely that ice can survive the age
of the solar system for an object with a radius of 0.3 km in the
orbit of P/2008 R1 for any plausible range of densities (0.5$<$$\rho$$<$1.3
gm cm$^{-3}$) \citep{Sarid12}.  However, some of the MBCs are members
of the Themis and and Beagle collisional families and have significantly
younger ages (~$\sim$2 Gyr and $\sim$10 Myr, respectively).  Even
small parent body fragments (down to $R_N$=0.3 km) are expected to
have surviving ice over a 10 Myr period.  Specifically for P/2008
R1 Garradd with its current size and orbit, ice can only survive
on the 10's of Myr timescale, and can be at depths as shallow as
10-15 m.

\citet{JYH09} have shown that the orbit of P/2008 R1
Garradd is not stable on timescales of 20-100 Myr due to its
interaction with the 8:3 mean-motion resonance with Jupiter.  This
is shorter than the likely collisional lifetime for objects this
size in the main belt.  The MBC 238P/Read was shown to have a
similarly small radius of $0.3\,\rm km$ \citep{2009AJ....137..157H},
and is one of the more active MBCs, and has now been seen active
at two perihelion passages.  Like P/2008 R1, 238P/Read is also not a member
of either the Themis or Beagle families.  It is likely that there are
more collisional families in the outer belt. It may be that objects
in the outer asteroid belt contain significant water-ice, but that the
objects are not discovered frequently because of their low level of
activity and the need for a collisional trigger.

\section{Conclusions} \label{sec:conclusions}

We undertook an intensive recovery effort of Main Belt Comet P/2008
R1 Garradd, in support of the {\sl Proteus} NASA Discovery mission
proposal.  Because it was significantly fainter than expected, all
our observations failed to recover the object, until we obtained
6000s of 8-m telescope time on Gemini North, combined with an
improved detection pipeline and artifact reduction techniques,
managed to recover the the object at a low $5\sigma$ level.

From the absolute magnitude $H_R\approx 19.6$, we infer that the
radius of MBC P/2008 R1 Garradd is $R_N$ = 0.3 km.  Given its
smaller perihelion distance and larger orbital eccentricity, this 
suggests that this body cannot have been at its present orbit
at its present size over the age of the solar system. From previous
findings that 238P/Read is also very small, we infer that the
survival of ice in small bodies may be commonplace.

\acknowledgements{The TNG data were obtained by G. P. Tozzi (Arcetri),
who kindly made them available to us.  This research used the
facilities of the Canadian Astronomy Data Centre operated by the
National Research Council of Canada with the support of the Canadian
Space Agency.  We used data from the ESO Science Archive Facility.
This material is based, in part, upon work supported by the National
Aeronautics and Space Administration through the NASA Astrobiology
Institute under Cooperative Agreement No. NNA09DA77A issued through
the Office of Space Science, and in part on NSF grant AST-1010059.
}

\newpage

%%%%%%%%%%%%%%%%%%%%%%%%%%%%%%%%%%%%%%%%%%%%%%%%%%%%%%%%%%%%%%%%
%%%% observations table
%%%%%%%%%%%%%%%%%%%%%%%%%%%%%%%%%%%%%%%%%%%%%%%%%%%%%%%%%%%%%%%%
\begin{table}
\begin{center}
\caption{Summary of the observations and magnitudes}
\label{tab:observations}
\begin{tabular}{ccccccccc}
%\hline
Epoch UT$^1$ & Telescope   & Obs$^2$ & Exp.$^3$ & Total$^4$ & $R^5$& $\Delta^6$& $\alpha^7$& Mag $^8$\\
\hline\hline 
2009-11-19.0 & TNG/DOLORES & GT &  2 &  1800 & 3.22 & 2.41 & 11.7 &$>23.8$ \\
2009-11-20.0 & TNG/DOLORES & GT & 34 &  6120 & 3.22 & 2.41 & 11.4 &$>24.4$ \\
2010-01-13.1 & NTT/EFOSC2  & OH & 10 &  3000 & 3.35 & 2.47 &  8.7 &$>22.6$ \\
2010-02-18.2 & UH/CCD      & JP & 30 & 17150 & 3.42 & 2.92 & 15.2 &$>25.3$ \\
2010-02-18.2 & UH/CCD$^{\dag}$ & JP & 22 & 9950 & 3.42 & 2.92 & 15.2 &$>24.9$ \\
2010-04-04.2 & GN/GMOS     & SM & 20 &  6000 & 3.50 & 3.65 & 15.9 &  26.1$\pm$0.2 \\
\hline
\end{tabular}
\end{center}
1: UT date and time of mid-exposure of the sequence; 
2: Observers: GT: G.~P. Tozzi; OH: O.~R. Hainaut; JP:
J.~Pittichov\'a;  SM: Gemini queue observing mode;
3,4: Number of exposures, and total exposure time [sec]; 
5,6: helio- and geocentric distances [AU]; 
7: solar phase angle [degrees]; 
8: R-band.
$^{\dag}$ Reduced UH dataset, removing the frames where the expected position
of P/2008 R1 was on the field star charge bleed.

\end{table}
%%%%%%%%%%%%%%%%%%%%%%%%%%%%%%%%%%%%%%%%%%%%%%%%%%%%%%%%%%%%%%%%

%%%%%%%%%%%%%%%%%%%%%%%%%%%%%%%%%%%%%%%%%%%%%%%%%%%%%%%%%%%%%%%%
% Astrometry Table
%%%%%%%%%%%%%%%%%%%%%%%%%%%%%%%%%%%%%%%%%%%%%%%%%%%%%%%%%%%%%%%%
\begin{table}
\begin{center}
\caption{Astrometry}
\label{tab:astrometry}
\begin{tabular}{lccrr}
Position$^{\dag}$ & $\alpha$ & $\delta$ & $\Delta\alpha$ [arcsec] & $\Delta\delta$ [arcsec]\\
\hline
Old orbit    & 05:55:33.39 & +35:14:31.0 &                 \\
Measurement  & 05:55:29.17 & +35:14:16.7 & -52.47 & -14.22 \\
New orbit    & 05:55:29.26 & +35:14:15.5 & -50.59 & -15.50 \\
\hline
\end{tabular}
\end{center}
$^{\dag}$All positions refer to 2010-Apr-13 05:45; $\Delta\alpha$ and
$\Delta\delta$ refer to the offset from the old orbit position.
\end{table}
%%%%%%%%%%%%%%%%%%%%%%%%%%%%%%%%%%%%%%%%%%%%%%%%%%%%%%%%%%%%%%%%

%%%%%%%%%%%%%%%%%%%%%%%%%%%%%%%%%%%%%%%%%%%%%%%%%%%%%%%%%%%%%%%%
%%% image sequence
%%%%%%%%%%%%%%%%%%%%%%%%%%%%%%%%%%%%%%%%%%%%%%%%%%%%%%%%%%%%%%%%
%%%%%%%%%%%%%%%%%%%%%%%%%%%%%%%%%%%%%%%%%%%%%%%%%%%%%%%%%%%%%%%%

%%%%%%%%%%%%%%%%%%%%%%%%%%%%%%%%%%%%%%%%%%%%%%%%%%%%%%%%%%%%%%%%
%%% finder chart
%%%%%%%%%%%%%%%%%%%%%%%%%%%%%%%%%%%%%%%%%%%%%%%%%%%%%%%%%%%%%%%%
%%%%%%%%%%%%%%%%%%%%%%%%%%%%%%%%%%%%%%%%%%%%%%%%%%%%%%%%%%%%%%%%

%%%%%%%%%%%%%%%%%%%%%%%%%%%%%%%%%%%%%%%%%%%%%%%%%%%%%%%%%%%%%%%%
%%% image processing technique 
%%%%%%%%%%%%%%%%%%%%%%%%%%%%%%%%%%%%%%%%%%%%%%%%%%%%%%%%%%%%%%%%
%%%%%%%%%%%%%%%%%%%%%%%%%%%%%%%%%%%%%%%%%%%%%%%%%%%%%%%%%%%%%%%%

%%%%%%%%%%%%%%%%%%%%%%%%%%%%%%%%%%%%%%%%%%%%%%%%%%%%%%%%%%%%%%%%
%%% pixel histogram
%%%%%%%%%%%%%%%%%%%%%%%%%%%%%%%%%%%%%%%%%%%%%%%%%%%%%%%%%%%%%%%%
%%%%%%%%%%%%%%%%%%%%%%%%%%%%%%%%%%%%%%%%%%%%%%%%%%%%%%%%%%%%%%%%

\bibliographystyle{aa} 
\bibliography{garraddrec}

\begin{thebibliography}{16}
\expandafter\ifx\csname natexlab\endcsname\relax\def\natexlab#1{#1}\fi

\bibitem[{{Allington-Smith} {et~al.}(2002){Allington-Smith}, {Murray},
  {Content}, {Dodsworth}, {Davies}, {Miller}, {Jorgensen}, {Hook}, {Crampton},
  \& {Murowinski}}]{2002PASP..114..892A}
{Allington-Smith}, J., {Murray}, G., {Content}, R., {et~al.} 2002, \pasp, 114,
  892

\bibitem[{{Buzzoni} {et~al.}(1984){Buzzoni}, {Delabre}, {Dekker}, {Dodorico},
  {Enard}, {Focardi}, {Gustafsson}, {Nees}, {Paureau}, \&
  {Reiss}}]{1984Msngr..38....9B}
{Buzzoni}, B., {Delabre}, B., {Dekker}, H., {et~al.} 1984, The Messenger, 38, 9

\bibitem[{{Encrenaz}(2008)}]{2008ARA&A..46...57E}
{Encrenaz}, T. 2008, \araa, 46, 57

\bibitem[{{Fukugita} {et~al.}(1996){Fukugita}, {Ichikawa}, {Gunn}, {Doi},
  {Shimasaku}, \& {Schneider}}]{Fukugita+96}
{Fukugita}, M., {Ichikawa}, T., {Gunn}, J.~E., {et~al.} 1996, \aj, 111, 1748

\bibitem[{{Garaud} \& {Lin}(2007)}]{2007ApJ...654..606G}
{Garaud}, P. \& {Lin}, D.~N.~C. 2007, \apj, 654, 606

\bibitem[{{Garradd} {et~al.}(2008){Garradd}, {McNaught}, {Meyer}, {Herald}, \&
  {Marsden}}]{2008MPEC....S...46G}
{Garradd}, G.~J., {McNaught}, R.~H., {Meyer}, M., {Herald}, D., \& {Marsden},
  B.~G. 2008, Minor Planet Electronic Circulars, 46

\bibitem[{{Hainaut} {et~al.}(2012){Hainaut}, {Boehnhardt}, \&
  {Protopapa}}]{HBP12}
{Hainaut}, O., {Boehnhardt}, H., \& {Protopapa}, S. 2012, \aap, in press

\bibitem[{{Hainaut} {et~al.}(2004){Hainaut}, {Delsanti}, {Meech}, \&
  {West}}]{Hai+04}
{Hainaut}, O.~R., {Delsanti}, A., {Meech}, K.~J., \& {West}, R.~M. 2004, \aap,
  417, 1159

\bibitem[{{Hook} {et~al.}(2004){Hook}, {J{\o}rgensen}, {Allington-Smith},
  {Davies}, {Metcalfe}, {Murowinski}, \& {Crampton}}]{2004PASP..116..425H}
{Hook}, I.~M., {J{\o}rgensen}, I., {Allington-Smith}, J.~R., {et~al.} 2004,
  \pasp, 116, 425

\bibitem[{{Hsieh} {et~al.}(2009){Hsieh}, {Jewitt}, \&
  {Ishiguro}}]{2009AJ....137..157H}
{Hsieh}, H.~H., {Jewitt}, D., \& {Ishiguro}, M. 2009, \aj, 137, 157

\bibitem[{{Jewitt}(2012)}]{2012AJ....143...66J}
{Jewitt}, D. 2012, \aj, 143, 66

\bibitem[{{Jewitt} {et~al.}(2009){Jewitt}, {Yang}, \& {Haghighipour}}]{JYH09}
{Jewitt}, D., {Yang}, B., \& {Haghighipour}, N. 2009, \aj, 137, 4313

\bibitem[{{J{\o}rgensen}(2009)}]{jorgensenGMOSCalib}
{J{\o}rgensen}, I. 2009, \pasa, 26, 17

\bibitem[{{Kleyna} \& {Meech}(2010)}]{MPC70591}
{Kleyna}, J. \& {Meech}, K. 2010, MPC, 70579, 22

\bibitem[{{Sarid} {et~al.}(2012){Sarid}, {Prialnik}, \& {Meech}}]{Sarid12}
{Sarid}, G., {Prialnik}, D., \& {Meech}, K. 2012, \mnras, submitted

\bibitem[{{Snodgrass} {et~al.}(2008){Snodgrass}, {Saviane}, {Monaco}, \&
  {Sinclaire}}]{2008Msngr.132...18S}
{Snodgrass}, C., {Saviane}, I., {Monaco}, L., \& {Sinclaire}, P. 2008, The
  Messenger, 132, 18

\end{thebibliography}

\end{document}